\documentclass[prl,twocolumn,superscriptaddress]{revtex4}

\usepackage{epsfig}

\newcommand{\ran}{\rangle}

\newcommand{\order}{O}

\begin{document}

\title{Classical simulability and the significance of modular exponentiation in Shor's algorithm}

\author{Nadav Yoran}\email{N.Yoran@bristol.ac.uk} \author{Anthony J. Short}
\affiliation{H.H.Wills Physics Laboratory, University of Bristol,
Tyndall Avenue, Bristol BS8 1TL, UK}


\begin{abstract}
We show that a classical algorithm efficiently simulating the
modular exponentiation circuit, for certain product state input and
with measurements in a general product state basis at the output,
can efficiently simulate Shor's factoring algorithm. This is done by
using the notion of the semi-classical Fourier transform due to
Griffith and Niu, and further discussed in the context of Shor's
algorithm by Browne.
\end{abstract}

\maketitle

\noindent The most celebrated quantum algorithm to date is undoubtedly Shor's factoring algorithm.
Distilling the crucial elements in this algorithm which allow for the (assumed) speed-up it
exhibits, may lead to a better understanding of the power of quantum computation in general. Shor's
algorithm has two main components: modular exponentiation, and the quantum Fourier transform (QFT).
Of the two the first is basically a classical circuit employing classical gates (manipulating
computational basis states). The only quantum aspect of this circuit is that it maintains quantum
coherence, i.e. it can act on a superposition of classical inputs and yields the corresponding
superposition of classical outputs. The QFT on the other hand uses gates that have no classical
equivalents, such as conditional phase and Hadamard gates, and is considered as the truly quantum
component of the algorithm. The QFT is a key component not only in Shor's algorithm but also in
several related quantum algorithms such as phase estimation and discrete logarithm.

Yet, it was recently demonstrated \cite{aharonov, us} that an approximate quantum fourier transform
(which can be used in Shor's algorithm~\cite{coppersmith}) can be efficiently classically simulated
using tensor contraction methods. This was shown for any product input state and product state
measurements on the output, and furthermore for a class of entangled states \cite{us} (as input or
as basis for output measurements). This seems to indicate that the computational power of Shor's
algorithm lies with the modular exponentiation circuit. Here, we demonstrate that this is indeed
the case by the following simple observation: {\it Any classical algorithm that can efficiently
simulate the circuit implementing modular exponentiation for general product input states and
product state measurements on the output, allows for an efficient simulation of the entire Shor
algorithm  on a classical computer.} In other words the power of Shor's algorithm lies in the
ability to implement the classical modular exponentiation operation on a certain product state
input and with measurements in a general product state basis.

The above result is also true for any circuit that can replace the
modular exponentiation in the overall quantum circuit for Shor's
algorithm (Fig. 1). For instance, the multiplication operations used
in the log-depth version of the (quantum part of) Shor's algorithm
due to Cleve and Watrous~\cite{cleve}.

A particular method for classically simulating quantum circuits, in
which general product input and output states are {\it
automatically} taken into account, is tensor contraction
\cite{markov, briegel}. Therefore, Our observation implies that any
tensor contraction scheme that efficiently simulates modular
exponentiation would be able to simulate Shor's factoring algorithm
efficiently.

Let us first describe in exact terms what we consider as a
simulation of a quantum circuit for product input states and product
state measurements. For a given quantum circuit let us denote a set
of (general) single qubit measurements on the output by $M_1,
\ldots,M_n$, their corresponding sets of possible outcomes by
$(m_{1}, \ldots ,m_{n})$ and some specific outcomes of these
measurements by $(r_{1}, \dots ,r_{n})$. We say that a quantum
circuit can be efficiently simulated by a classical computer for
product state input and product state measurements if for any such
set of single qubit measurements and any product state input there
is an efficient classical algorithm for calculating the conditional
probabilities:
\begin{equation}
P(m_{i}|r_{j_{1}}, \ldots, r_{j_{k}})
\end{equation}
where the indices $j_{1}, \ldots ,j_{k}$ correspond to any subset of the measurements $(M_1, \ldots
,M_n)$ including the empty set. (Of course, there are exponentially many such conditional
probabilities, one for each value of the bit-string $r_{j_{1}}, \ldots, r_{j_{k}}$ therefore there
is no way to efficiently calculate all of them; however, we only require that each particular
conditional probability can be calculated efficiently.) Sampling from these conditional
probabilities qubit by qubit one is able, using the classical algorithm, to obtain a final outcome
with the same probability as it would have been obtained by the quantum computer.

Our definition above for a simulation of a quantum circuit is similar to the `density computation
of quantum circuit' given by Terhal and DiVincenzo~\cite{ter-div}. The difference being the fact
that here we allow general product state input and product state measurements at the output,
whereas in the weaker density computation only computational basis input and measurements in the
computational basis at the output are permitted.

The quantum part of Shor's algorithm, where we wish to factor an $n$
qubit integer $N$, is composed of the following steps (Fig. 1).
\begin{enumerate}
\item Initialize two registers, the first with $2n$ qubits and the
second with $n$ qubits, in the state  $|0\ran_{1}|0\ran_{2}$.

\item Apply a Hadamard gate on each of the qubits of register $1$.
the state of the computer would now be:
\begin{equation}
 (|+\ran \cdots |+ \ran)_{1}|0\ran_{2}\ =
 \sum_{x=0}^{2^{2n}-1}|x\ran_{1}|0\ran_{2} \; , \label{2}
\end{equation}
where $|+\ran = 1/\sqrt{2}(|0\ran+|1\ran)$.

\item Apply modular exponentiation. Namely, apply the
unitary operation:
\begin{equation}
 |x\ran_{1}|0\ran_{2} \rightarrow
 |x\ran_{1}|a^{x}\,\mbox{mod}\,N\ran_{2} \; ,
\end{equation}
where $a$ is a randomly chosen number ($a<N$) co-prime with $N$.

\item Apply a QFT on the first register.

\item Measure the first register in the computational basis.

\end{enumerate}

\begin{figure}\begin{center}
\epsfig{file=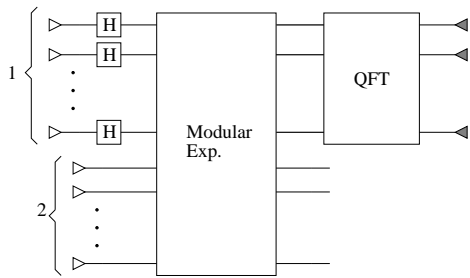} \caption{The quantum circuit for Shor's
algorithm. The empty triangles on the left side represent the
computational basis input state the shaded triangles on the right
represent measurements in the computational basis and the boxes
denoted by H stand for Hadamard gates. The QFT together with the
output measurements can be replaced by the semi-classical circuit in
Fig. 2.\label{fig:shor} }
\end{center}\end{figure}

In order to prove our result above we make use of the following fact, first demonstrated by
Griffith and Niu~\cite{grn}: The QFT circuit followed by measurements in the computational basis
can always be replaced by a `semi-classical' QFT circuit which includes only single qubit gates,
measurements in the computational basis, and feed-forward (without any two qubit gates). Griffith
and Niu observed that when a controlled unitary is immediately followed by a measurement in the
computational basis on the control qubit, the gate operation can be implemented by first measuring
the control qubit and then applying a gate on the second qubit according to the outcome (that is,
the gate would only be applied if the control is measured in the required state). In Shor's
algorithm the QFT is immediately followed by measurements in the computational basis
(see~\cite{nielsen} for example) therefore it can be replaced by the semi-classical circuit (shown
in Fig. 2).

\begin{figure}\begin{center}
\epsfig{file=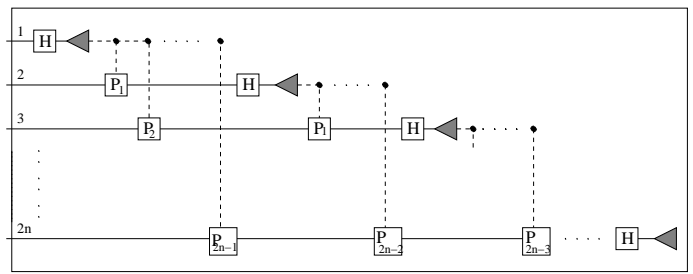} \caption{The semi-classical QFT circuit.
The triangles denote measurements in the computational basis and the
dashed lines signify the fact that the application of the different
single qubit phase gates (denoted by $P_{i}$) is conditioned on the
outcomes of those measurements. \label{fig:semi} }
\end{center}\end{figure}

Browne \cite{browne} used the semi-classical QFT to show that in the
case where the modular exponentiation circuit together with Hadamard
gates at the input does not produce much entanglement, and therefore
can be simulated classically \cite{vidal,jozsa}, Shor's algorithm
can be efficiently simulated as well (typically modular
exponentiation does produce highly entangled states \cite{jozsa},
however there may be special cases for which it does not).

Let us now assume that the modular exponentiation circuit can be
simulated efficiently when the input is a direct product and the
output is subjected to single qubit measurements. Our simulation of
Shor's algorithm proceeds via an iterative procedure as follows:

\begin{enumerate}
\item First, we calculate the probabilities $P_{Shor}(m_{1})$ for a
measurement on the first qubit (in the first register) in Shor's
algorithm.

From our assumption it follows that these probabilities can be
efficiently calculated. Indeed, in the Shor circuit with the
semi-classical QFT (which obviously produces the same output as the
original Shor algorithm) the measurement of the first qubit in the
computational basis at the output of the semi-classical QFT is
nothing else than a single qubit measurement of the output of the
modular exponentiation. (More precisely, it is the measurement of
the first qubit of the output of the modular exponentiation in the
Hadamard transformed basis.) Furthermore, the input state into the
modular exponentiation is a direct product state (the state $(|+\ran
\cdots |+ \ran)_{1}|0\ran_{2}\,$) so the conditions in our
assumption apply.

\item We sample from the distribution $P_{Shor}(m_{1})$. Let $r_1$ be the result of this sampling.

\item We calculate the conditional probabilities
$P_{Shor}(m_{2}|r_1)$ for the measurement on the second qubit in
Shor's algorithm, given the outcome $r_1$ for the first qubit.

Again, from our assumption it follows that these probabilities can
be efficiently calculated. Once the output $r_1$ of the first qubit
is fixed, we know what is the feed-forward from the first qubit in
the semi-classical QFT. With this knowledge, the measurement in the
computational basis of the second qubit in the circuit for Shor's
algorithm becomes a well-defined measurement on the second qubit of
the output of the modular exponentiation circuit.

\item We repeat steps (2) and (3) for $i=2,\ldots 2n$. That is in
the next step we sample from the conditional probability distribution $P_{Shor}(m_{2}|r_1)$ and
obtain outcome $r_{2}$, and in general we calculate and then sample from
\begin{equation}
 P_{Shor}(m_{i}|r_{1},\cdots,r_{i-1}) \quad \mbox{for}\; 2\leq i\leq
 2n \;,
\end{equation}
where the basis for the measurement $m_{i}$ is set according to the
outcomes of previous measurements $r_{1},\ldots,r_{i-1}$.

\end{enumerate}
At the end of the process an outcome ($r_{1}, \ldots,r_{2n}$) is
obtained with the same probability it would have been obtained by
measuring the output of the quantum circuit implementing Shor's
algorithm.

Clearly, for the purpose of simulating the Shor algorithm it is
enough to consider only one input state (Eq.~\ref{2}). (Note that we
could not simply redefine this as a new computational basis state
$|0^{\prime} \ran$ if we do not want to change the modular
exponentiation circuit.) For the output measurements, however, we
need to take into account every possible phase gate. It was noted by
Browne \cite{browne} that if we consider the above input state but
allow output measurements only in the computational basis, then the
modular exponentiation circuit can be simulated efficiently. In the
same way it is not hard to see that if one considers only
computational basis input states and allows any product state
measurement at the output, then the circuit would be simulable as
well. Thus, in some sense, our requirements are the minimal ones
with which quantum advantage is achieved.

Consider now a tensor contraction method for simulation \cite{markov, briegel}. In these methods
one associates tensors to circuit elements -- one and two qubit gates, single-qubit input and
output states. The latter may correspond to outcomes of single qubit measurements or to unmeasured
output qubits. The rank of the tensors is determined by the number of (input and output) qubits on
which the circuit element operates. That is, the tensor has an index for each input or output wire
connected to the circuit element. Thus, for an input state or an output measurement correspond to
tensors of rank one, single-qubit gates correspond to rank two tensors and two-qubit gates are
represented by rank four tensors. Two circuit elements connected by a qubit wire (the output of one
is the input of the other) share a joint index. A probability for obtaining a certain outcome at
one or more output qubits can be calculated by contracting (summing over all indices of) all
tensors representing the circuit with the appropriate configuration at the outputs (tensors
corresponding to the required outcomes for the measured qubits and tensors corresponding to
unmeasured qubits for the rest). The problem with such a contraction process is that the number of
terms is exponentially large. To avoid this, the tensors are contracted one at a time breaking the
overall sum to a series of separate sums, where in each step two existing tensors are replaced with
a new tensor obtained by summing over joint indices. For instance summing over a joint index of a
pair of two qubit gates connected by a single qubit wire one obtains a tensor of rank six (e.g.
$T^{kl}_{ij} T^{no}_{lm} \rightarrow T^{kno}_{ijm}$). A tensor contraction simulation is efficient
(i.e. can be implemented in polynomial time) if the tensors generated in the procedure have at most
$\order(\log n)$ indices.

The only factor which determines the complexity of a contraction process of a given quantum circuit
is its topology. The type of gates or their actual operation on the input (beyond the fact that it
is linear) is irrelevant. furthermore rank one tensors, representing the input and output elements,
and rank two tensors corresponding to single-qubit gates, do not affect the topology of the circuit
and therefore do not affect the efficiency of the simulation. For example, these tensors can be
incorporated into tensors of neighbouring elements by contracting them together at the first stage
of the simulation. This produces new tensors with the same number of indices as the original
neighbouring tensor (in the case of one-qubit gates) or less (in the case of input or output
elements) without changing the graph of connections of the circuit.

From our discussion above it is clear that a tensor contraction
simulation for a given quantum circuit with a certain set of input
and output elements would also work for a any other set of input
states and output measurements as long as these are single qubit
states and measurements. Thus, a tensor contraction algorithm
simulating the modular exponentiation would also be able to simulate
Shor's factoring algorithm. Note that this implies that it is
unlikely that a tensor contraction scheme would be able to
efficiently compute modular exponentiation even for 'classical'
computational basis input and output states, a task which obviously
can be done by other classical algorithms.

So far we have discussed only modular exponentiation. However from our method of simulation it is
clear that any quantum circuit with the same structure (as in Fig. 1), where the modular
exponentiation is replaced by some other unitary operation $U$, can be simulated efficiently on a
classical computer if $U$ is efficiently simulable for product state input states and measurements,
and in particular if $U$ has an efficient tensor contraction scheme.

\smallskip
\acknowledgments

The authors wish to thank S. Popescu, D. E. Browne and D. J.
Shepherd for fruitful discussions. The work of N.~Y. was supported
by UK EPSRC grant (GR/527405/01), and A.J.S. was supported by the UK
EPSRC's ``QIP IRC'' project.

\end{document}